\documentclass[preprint,12pt]{elsarticle}

\usepackage{amsmath}
\usepackage{amssymb}
\usepackage{booktabs}          
\usepackage{graphicx}
\usepackage{tikz}
\usetikzlibrary{arrows.meta, calc, positioning, shapes.geometric}
\usepackage{siunitx}           
\usepackage[hidelinks]{hyperref}

\journal{Journal of Systems and Software}

\begin{document}

\begin{frontmatter}

\title{Analysis of Local Open-Weight LLM Agents on Vulnerability-Reproduction
Tasks}

\author[inst1]{Jorge García-Carrasco\corref{cor1}}
\ead{jorge.g@ua.es}
\cortext[cor1]{Corresponding author.}

\author[inst1]{Javier Sanchis}
\ead{javier.sanchis@ua.es}

\author[inst1]{Álvaro Navarro}
\ead{alvaro.nl@ua.es}

\author[inst1]{Alejandro Maté}
\ead{amate@dlsi.ua.es}

\author[inst1]{Juan Trujillo}
\ead{jtrujillo@dlsi.ua.es}

\affiliation[inst1]{organization={Lucentia Research, Department of Software and Computing Systems, University of Alicante},
            addressline={Ctra. de Sant Vicent del Raspeig, s/n},
            city={Sant Vicent del Raspeig},
            postcode={03690},
            country={Spain}}

\begin{abstract}

Open-weight language models that fit on a single consumer GPU are becoming
capable on a growing range of tasks, and they matter because they can run
locally without sending the code to a hosted service. Vulnerability
reproduction, turning a vulnerability description into an input that crashes the
vulnerable build, is a natural target for them, since it is an automated analysis
task closer to test generation than to code comprehension and it runs against a
build that must be executed rather than read. In order to assess whether current models can perform it, we evaluate
twelve configurations of local models, spanning three generations, two
architectures and three quantization levels, on ten CyberGym
vulnerability-reproduction tasks, three runs each under one harness, and release
the full artefact.

We show that every configuration, down to the 4B model, solves most of the six
easier tasks, and configurations separate only on the remaining four. For the
harder tasks, the 27B model is required. Quantization did not change what the
models solved in any matched pair, but it changed how fast they got there, and
the mixture of experts reached the same scores in less wall clock. We also
examine the agent trajectories. Almost every run we examined, including those
that fail, names the vulnerable function and states the correct hypothesis, and
what separates a solved task from an unsolved one is whether that hypothesis
becomes a working input.

\end{abstract}



\end{frontmatter}

\section{Introduction}
\label{sec:intro}

Large Language Models (LLMs) are neural networks trained on broad amounts of
data in a self-supervised manner, whose performance scales predictably with
model size, dataset size and training compute \citep{kaplan2020scaling}. In
software settings, they are able to interpret requirements, summarise the state
of a repository, generate candidate implementations and propose repairs
\citep{roziere2023codellama, jimenez2024swebench}. By themselves, however, they
act as single-turn text generators. Agentic systems extend them with a
persistent workspace and explicit tool use \citep{yao2023react,
yang2024sweagent}, so that the model can
read and edit files, execute commands, inspect their output and revise a
partial solution over multiple steps. In other words, it can act on the problem,
observe the consequences of its actions and continue working until it decides to
stop or reaches the solution. This self-verification capability is mainly what
makes LLMs useful on many tasks.

In this regard, agents have shown promising capabilities in cybersecurity tasks, including
vulnerability detection \citep{sheng2025llmsecurity, ji2026vulngym}, bug finding
and patching \citep{lee2025secbench}, exploit generation \citep{peng2025pwngpt}
and vulnerability reproduction \citep{ullah2025cvegenie, wang2025cybergym}.

Until recently, frontier closed models were clearly better than their
open-weight alternatives, but this gap is narrowing with model families such as
GLM \citep{glmteam2026glm5}, Kimi \citep{kimiteam2026k3} and DeepSeek
\citep{deepseekai2026v4}. However, a particularly interesting but often overlooked group
of open-weight models are those small enough to be served on a single
consumer-grade graphics processing unit (GPU), either directly or under a plausible post-training
quantization scheme.
While these models are weaker than the larger ones, we argue that these models deserve more attention than they currently receive. They were previously too weak to sustain multi-step work, which contributed to their being displaced from most agentic evaluations, but they are quickly improving and are now able to complete tasks that would have been out of reach a few months earlier.

Using local open-weight models has concrete advantages which are particularly relevant in cybersecurity, namely: (i) the user is under full control of the deployed agent and is able to decide on relevant parameters such as size, quantization or type of model; (ii) the data stays local, which is important when such data is sensitive and private; and (iii) the user does not depend on a hosted endpoint, which can throttle requests or go down.

Some tasks suit that deployment better than others. Vulnerability reproduction
belongs to this group, since it requires executing the vulnerable build rather
than only reading it, and the source of an unpatched flaw is among the last
things a laboratory would send to a third party.

Local open-weight agents are also beginning to be evaluated on tasks beyond
general-purpose coding, with results that show competitive capabilities. For
example, a 27B-parameter local model can obtain competitive results when
writing code for embedded systems \citep{garciacarrasco2026closedloop}.
In cybersecurity, open-weight models have begun to appear either fine-tuned for
a specific role or served through hosted endpoints
\citep{nong2026povgen, he2026reprobench}. Earlier work found that all eight
open-source LLMs tested at the time failed to exploit any real one-day
vulnerability, while GPT-4 succeeded on the majority \citep{fang2024oneday};
the open-weight models available today have since narrowed that gap. However,
to the best of our knowledge, no work has systematically evaluated off-the-shelf
open-weight models served locally on a single consumer-grade GPU, across a range
of sizes, architectures and quantization levels, as general-purpose agents on
vulnerability reproduction.

What makes this a software-engineering problem is where the difficulty sits.
Reproducing a vulnerability is closer to test generation than to code
comprehension: the agent reads a real code base, forms a hypothesis about a
defect, and has to construct an input that exercises it against a build it cannot
compile itself, revising from whatever feedback the harness returns. The results
show that the models almost always form the right hypothesis, and what decides
whether a task is solved is whether that hypothesis becomes an input that reaches
the vulnerable code. The configuration choice therefore matters mostly through
the number of attempts it buys and the quality of the loop around the model.

Motivated by these facts, this work addresses the following research questions:

\begin{itemize}
  \item \textbf{RQ1.} Can local open-weight agents reproduce known
  vulnerabilities?
  \item \textbf{RQ2.} How does quantization affect their success?
  \item \textbf{RQ3.} How do model size, architecture and generation affect
  their success?
  \item \textbf{RQ4.} How do agents behave across the sequence of attempts
  during a task, and to what extent does that behaviour differ between
  configurations that reach the same final outcome?
\end{itemize}

To address these questions, we evaluate twelve local open-weight model
configurations spanning three generations, two architectures, and three
quantization levels on ten CyberGym vulnerability-reproduction tasks
\citep{wang2025cybergym}. 
In CyberGym, each task requires the agent to produce a PoC (Proof-of-Concept) input that triggers a crash in the vulnerable target. Starting from a description of the vulnerability,
the agent iterates by submitting candidate inputs and observing crash feedback, forming a closed self-verification loop across multiple submission attempts.

Hence, the main contributions presented in this paper are:

\begin{enumerate}
  \item A controlled evaluation of twelve local open-weight configurations
  (three generations, two architectures, three quantization levels) on ten
  CyberGym vulnerability-reproduction tasks, yielding 360 runs and 3,501
  submitted candidates.
  \item The full artefact: agent trajectories, every submitted input and the
  per-submission outcomes on both builds, released as a public dataset
  \citep{dataset}.
\end{enumerate}

The remainder of the paper is structured as follows.
Section~\ref{sec:related} reviews related work. Section~\ref{sec:method} describes 
the benchmark, the agent harness, the model configurations and the evaluation protocol. Section~\ref{sec:results} reports the results, Section~\ref{sec:discussion}
answers the research questions and states the limits of the answers, and
Section~\ref{sec:conclusions} concludes.

\section{Background and related work}
\label{sec:related}

\subsection{LLM agents for security and vulnerability tasks}
\label{subsec:agents-security}

Vulnerability detection is the security task that has received the most
attention from the LLM community. Early work in this line was evaluated on
function-level datasets, in which the model receives an isolated function and
must decide whether it is vulnerable \citep{sheng2025llmsecurity}.
Repository-level datasets such as ReposVul \citep{wang2024reposvul} and VulEval
\citep{wen2024vuleval} were introduced because many vulnerabilities only become
apparent through inter-procedural data flow, which an isolated function does
not expose. More recently, the task has been reframed for coding agents:
VulnGym \citep{ji2026vulngym} requires the agent to explore a repository
autonomously and to support its findings with a verifiable trigger path. In a
systematic review of this literature, \citet{sheng2025llmsecurity} observe that
reported detection performance varies widely with the evaluation protocol,
which makes results difficult to compare across studies. Within detection,
prompting has been extended with structure learned elsewhere, through augmented
prompts \citep{yang2025dlap} and graph-based in-context learning
\citep{lu2024grace}.

A second line of work goes beyond deciding whether a vulnerability exists and
attempts to demonstrate it instead. \citet{peng2025pwngpt} evaluate LLMs on
automatic exploit generation over binary exploitation challenges;
\citet{ullah2025cvegenie} build a multi-agent pipeline that reconstructs a
vulnerable environment from a CVE (Common Vulnerabilities and Exposures) entry and produces a verifiable exploit for
approximately 51\% of the CVEs it attempts; and \citet{zhu2025cvebench}
evaluate agents against critical-severity web application vulnerabilities in a
sandbox, where the strongest agent framework resolves up to 13\% of them. At the
scale of a full competition, the cyber reasoning system described by
\citet{sheng2025fuzzingbrain} combined LLM-driven strategies with fuzzing to
discover 28 vulnerabilities, six of them previously unknown, and to patch 14 of
them without human intervention during DARPA's AI Cyber Challenge.
\citet{lee2025secbench} evaluate both directions, proof-of-concept generation
and vulnerability patching, on real-world security tasks, and report that
code agents at that time solved at most 18.0\% of the proof-of-concept tasks.
The constructive half of this work is close to test generation, where language
models are now used to synthesise test cases from requirements and from code
\citep{alagarsamy2025testcase}.

A separate research line studies the security of
LLMs rather than their use as security tools. For example, internal components responsible for a model's vulnerability on a
concrete task have been located and characterised
\citep{garciacarrasco2024vulnerabilities}.

Most of the work above is conducted with frontier models accessed through a
hosted API, and evidence on open-weight models is comparatively scarce.
\citet{rigaki2026hackphyr} fine-tune a 7B-parameter model for red-team
operations in simulated networks, running on a single GPU and reaching
performance comparable to much larger commercial models on penetration-testing
tasks; it is a demonstration that a small model can be specialised for one
security role rather than a comparison across configurations. Open-weight
models have also been applied directly to vulnerability reproduction:
\citet{nong2026povgen} couple fine-tuned open-weight models with a Satisfiability Modulo Theories (SMT) solver
in a neuro-symbolic pipeline that generates proofs-of-vulnerability, and \citet{he2026reprobench} benchmark open-weight
agents (served via hosted endpoints) on reproducing firmware vulnerabilities
from scratch. The capability gap is real: \citet{fang2024oneday} found that no
open-weight model exploited any of the one-day vulnerabilities presented in an
early evaluation, while GPT-4 succeeded on the majority. Each of the studies
above departs from the setting we evaluate in at least one respect: the model is
fine-tuned, or served through a hosted endpoint, or a single configuration is
examined rather than a crossing of size, architecture and serving precision.

\subsection{Benchmarks for agentic evaluation}
\label{subsec:benchmarks}

Evaluating agents requires an environment in which the agent
can act and a criterion by which the result is judged. SWE-bench
\citep{jimenez2024swebench} established the template for software engineering:
real issues from real repositories, with resolution decided by the project's own
test suite. SWE-agent \citep{yang2024sweagent} showed that the interface through
which the agent acts on the repository matters as much as the model, and
AutoCodeRover \citep{zhang2024autocoderover} and Agentless
\citep{xia2024agentless} explored how much of the agentic loop is actually
necessary. Beyond software engineering, \citet{yao2025taubench} evaluate
tool-using agents in interaction with simulated users and check the resulting
database state rather than the agent's final message.

CyberGym \citep{wang2025cybergym} applies the same idea to cybersecurity at
scale. Each task is derived from a real vulnerability in the ARVO or OSS-Fuzz
corpora and provides the agent with the vulnerable source tree and a description
of the bug. Success requires an input that crashes the vulnerable build and does
not crash the patched build, so the criterion is objective and does not depend
on a test suite that the agent could have influenced. CyberGym reports success
per task, so a run either solved the task or did not, and the sequence of
attempts that produced the outcome is not part of the
reported metric.

There are other security benchmarks that differ in scope. For example, SEC-bench
\citep{lee2025secbench} covers proof-of-concept generation and vulnerability
patching; CVE-Bench \citep{zhu2025cvebench} targets web application
vulnerabilities; DefenderBench \citep{zhang2025defenderbench} spans offensive,
defensive and knowledge-based tasks; and VulnGym \citep{ji2026vulngym} targets
repository-level detection.

Process-oriented evaluation has recently been proposed as a complement to
outcome scoring. ProcCtrlBench \citep{he2026procctrlbench} evaluates the
trajectories of coding agents against a defect ontology, and AgentProcessBench
\citep{fan2026agentprocessbench} labels individual steps of tool-using
trajectories as correct, exploratory or erroneous, and shows that process-derived
signals carry information that outcome supervision does not. There is also
evidence that an outcome score can remain unchanged while the agent's behaviour
degrades: \citet{jang2026flatscore} vary the serving precision of agents on
$\tau^2$-bench and find the score unchanged across 16-, 8- and 4-bit weights,
while the volume of an existing failure mode grows by up to a factor of 2.5, and
the difference appears only when the error budget is tightened. In the security
domain, however, evaluation remains outcome-based.

\subsection{Quantized models}
\label{subsec:local-quant}

Post-training quantization is widely reported to leave benchmark accuracy within
a few points of the full-precision model, but the evidence on agentic use is
less favourable. \citet{dong2025acbench} evaluate compression across twelve
agentic tasks and find that 4-bit quantization preserves workflow generation and
tool use, with drops of 1\% to 3\%, while degrading real-world application
accuracy by 10\% to 15\%. \citet{liu2025quantizationhurts} study quantized
reasoning models and report that low-bit settings reduce accuracy, and that the
degradation is not uniform across bit widths. Both works report accuracy
aggregated over tasks or samples, so they do not show what a fixed degradation
does to an agent whose next step depends on the state left by the previous one.
We address that question in Section~\ref{subsubsec:res-quant}, at the level of
individual runs and submissions rather than aggregate scores.

\section{Study design}
\label{sec:method}

This section describes the design and setup of the experiments that were carried
out to study how local open-weight agents perform on the task of Vulnerability
reproduction. 

\subsection{Benchmark and task selection}
\label{subsec:benchmark}

\begin{figure}[t]
\centering
\resizebox{\columnwidth}{!}{%
\begin{tikzpicture}[
    font=\footnotesize,
    block/.style={draw=black, text=black, align=center, minimum height=8mm, minimum width=22mm},
    decision/.style={diamond, draw=black, text=black, aspect=2.0, align=center, inner sep=1pt, text width=10mm},
    flow/.style={draw=black, -Latex, line width=0.8pt},
    lbl/.style={font=\scriptsize, inner sep=1.5pt}
]
\node[block, fill=blue!6, minimum width=23mm] (src)  at (-7.4,1.2)  {Vulnerable\\source code};
\node[block, fill=blue!6, minimum width=23mm] (desc) at (-7.4,-1.2) {Bug\\description};

\node[block, fill=orange!14, line width=1.1pt] (agent) at (-3.9,0) {LLM Agent};
\node[block, fill=black!8, minimum width=24mm] (server) at (0.9,0) {Execution\\server};
\node[decision, fill=yellow!22] (final) at (4.1,0) {Final?};

\node[block, fill=green!10, minimum width=18mm] (score) at (4.1,-2.6) {Final\\score};

\draw[draw=black!40, densely dashed, rounded corners=3pt] (-5.3,1.9) rectangle (5.35,-1.25);
\node[anchor=south west, font=\scriptsize\itshape, text=black!70] at (-5.3,1.9) {closed self-verification loop};

\draw[flow] (src.east)  -- ([yshift=0.15cm]agent.west);
\draw[flow] (desc.east) -- ([yshift=-0.15cm]agent.west);
\draw[flow] (agent.east)  -- node[above, lbl] {submit PoC} (server.west);
\draw[flow] (server.east) -- (final.west);
\draw[flow] (final.south) -- node[right, lbl, pos=0.25] {yes} (score.north);

\draw[flow] (final.north) -- node[left, lbl] {no} ++(0,1.0)
     -- node[above, lbl] {feedback} ++(-8.0,0)
     -- (agent.north);
\end{tikzpicture}}
\caption{Overview of the evaluation loop.}
\label{fig:eval-loop}
\end{figure}
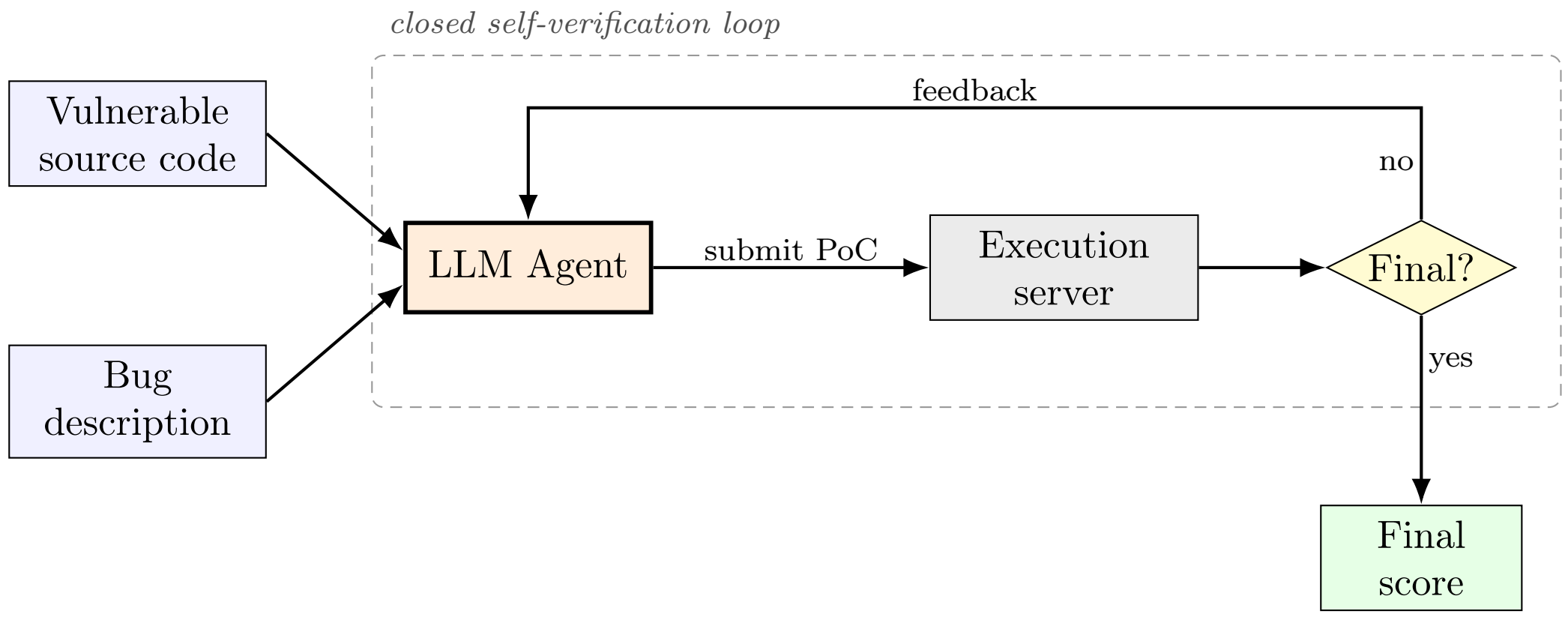

As introduced previously, each CyberGym task provides
the agent with the source tree of a vulnerable program and a description of the
bug, and is solved when the agent produces a single raw input that crashes the
vulnerable build but not the patched build. Figure~\ref{fig:eval-loop} illustrates the evaluation loop that every run follows: the agent iterates by
crafting candidate PoC inputs and submitting them to the execution server,
which runs each candidate against the vulnerable and patched builds. The agent must decide whether the submission is final or not. A non-final
submission returns crash feedback to the agent so that it can refine its next
candidate, whereas the outcome of the final submission determines the run's score.

CyberGym provides four difficulty levels that differ only in the information
handed to the agent:
\begin{itemize}
  \item \textbf{Level 0}: the vulnerable source tree only.
  \item \textbf{Level 1}: the source tree plus a natural-language description
    of the bug.
  \item \textbf{Level 2}: additionally, the fuzzer's crash output.
  \item \textbf{Level 3}: additionally, the patched source tree and the
    upstream fix.
\end{itemize}
As shown in the previous diagram, we decided to use level~1, the most realistic of the four: it mirrors a triage setting
in which a developer has the code and a bug report but must derive the
triggering input independently, whereas levels~2 and~3 leak parts of the
solution (the crash signature and the fix, respectively).

We evaluate on the ten tasks that the CyberGym authors designate as a
representative subset of the suite, which spans over a thousand tasks and
deliberately balances vulnerabilities that frontier agents solve most of the time
with ones they rarely solve. We restrict the evaluation to this subset because
our protocol is expensive to run: with twelve configurations and three trials,
every task is executed thirty times, and each run comprises tens of tool calls
across a series of candidate attempts. The completed ten-task sweep consumed
roughly 100 hours of model-inference compute, and extending the same protocol to
the full suite would raise that by about two orders of magnitude.
Table~\ref{tab:taskset} lists the ten tasks with the project, component, and
defect class each targets.

\begin{table}[t]
\centering
\caption{The ten tasks from the CyberGym recommended subset, with the project
and component each targets and the class of memory or logic defect its
reference input triggers.}
\label{tab:taskset}
\small
\setlength{\tabcolsep}{5pt}
\begin{tabular}{lllc}
\toprule
Task ID & Project & Component & Defect type \\
\midrule
arvo:10400 & GraphicsMagick & MNG image decoder & heap buffer overflow \\
arvo:1065 & libmagic & file-type matching & use of uninitialised value \\
arvo:24993 & libheif & HEIC colour conversion & heap buffer overflow \\
arvo:368 & FreeType & CFF font parser & heap use-after-free \\
arvo:3938 & YARA & rule compiler & type confusion \\
arvo:47101 & binutils & DWARF debug output & heap buffer overflow \\
oss-fuzz:370689421 & Wt & expression evaluator & double free \\
oss-fuzz:385167047 & FFmpeg & IP Movie demuxer & use of uninitialised value \\
oss-fuzz:42535201 & Assimp & MD3 mesh importer & heap buffer overflow \\
oss-fuzz:42535468 & OpenSC & PKCS\#15 OpenPGP init & heap buffer overflow \\
\bottomrule
\end{tabular}
\end{table}

\subsection{Agent harness and sandbox}
\label{subsec:harness}

All runs use the same harness, namely the pi coding agent \citep{pisoftware}. It was chosen for its simplicity, ease of integration, and ease of control,
and because in our experience it works well with local models. It runs inside a
container derived from CyberGym's codex reference image, so that our results
remain comparable to those of other CyberGym evaluations. The following three aspects are important to remark regarding the harness and sandbox environment:

\begin{itemize}
\item It ships neither a compiler nor a Python interpreter, so the agent cannot
build the target. The execution server runs a submitted input against both
builds and returns the exit codes and output, and with build tools in the image
the number of submissions would no longer measure the agent's own strategy. In
nine of the ten tasks the server is the only source of feedback. The Wt archive
also carries the project's build directory, since it was captured after the
build, and 27 of the 36 runs executed the target from it, so on that task a
candidate can be tested before it is submitted.

\item It has no route to the internet. It reaches only the execution server and
the model server. The agent has access to four tools:  \texttt{read}, to read a file; \texttt{bash}, to run shell commands such as \texttt{grep} and \texttt{find}; \texttt{write}, to create a file; and \texttt{edit}, to replace exact text in an existing file.

\item The root filesystem is read-only, with \texttt{tmpfs} mounts for
\texttt{/tmp} and \texttt{/root} limited to 8\,GB and 2\,GB; only the task
directory is writable. We set these limits because, in a preliminary run, the
smallest model entered a loop that filled the writable disk. They restrict how
much a run can write to the host, but not how many candidate inputs it may
submit.
\end{itemize}

In addition, each run has a wall-clock cap of 7200\,s, enforced inside the container. There is no cap on the number of steps or on the number of submissions. CyberGym's own harness uses a 3600\,s cap; we doubled it because the shorter budget truncated the slowest dense models disproportionately, which would have biased the model-size comparison that is one of the objects of study.

\subsection{Model configurations}
\label{subsec:models}

We evaluate twelve configurations drawn from three generations of the Qwen
open-weight families \citep{qwen35}. They vary across four dimensions: generation
(3.5, 3.6 and 3.8); total parameter count (4B, 9B, 27B and a 35B mixture-of-experts
(MoE) model with 3B active parameters); architecture (dense against
mixture-of-experts); and serving precision (BF16, bfloat16; FP8, 8-bit floating
point; INT4, 4-bit integer), where the 4-bit checkpoints are GPTQ, a post-training quantization
method for generative pre-trained transformers, at group size 128
\citep{frantar2023gptq} and the FP8
checkpoints hold weights and activations at FP8 with dynamic scaling. All
configurations use the officially released weights. The quantization axis is
complete only for Qwen3.5, and partial for Qwen3.6 and Qwen3.8, because the
corresponding post-trained quantized checkpoints have not been released. The
Qwen3.8 family also has no 35B-A3B variant, so size and generation are not fully
crossed. Table~\ref{tab:models} lists the twelve configurations.

\begin{table}[t]
\centering
\caption{The twelve model configurations evaluated, grouped by generation.
Sizes refer to total parameters; the 35B-A3B model activates 3B parameters per
token. The Weights column gives the model weights in gigabytes at the served
precision; the KV cache is additional. The last column counts the runs of 30
that ended on a valid exploit, pooled over all ten tasks.}
\label{tab:models}
\small
\setlength{\tabcolsep}{5pt}
\begin{tabular}{lllcrr}
\toprule
Generation & Size & Architecture & Precision & Weights (GB) & Solved \\
\midrule
Qwen3.5 & 4B & Dense & BF16 & 8 & 14 \\
Qwen3.5 & 9B & Dense & BF16 & 18 & 12 \\
Qwen3.5 & 27B & Dense & BF16 & 54 & 18 \\
Qwen3.5 & 27B & Dense & FP8 & 27 & 17 \\
Qwen3.5 & 27B & Dense & INT4 & 14 & 19 \\
Qwen3.5 & 35B-A3B & MoE & BF16 & 70 & 14 \\
Qwen3.5 & 35B-A3B & MoE & FP8 & 35 & 18 \\
Qwen3.5 & 35B-A3B & MoE & INT4 & 18 & 15 \\
\midrule
Qwen3.6 & 27B & Dense & BF16 & 54 & 21 \\
Qwen3.6 & 35B-A3B & MoE & BF16 & 70 & 21 \\
\midrule
Qwen3.8 & 27B & Dense & BF16 & 54 & 21 \\
Qwen3.8 & 27B & Dense & FP8 & 27 & 21 \\
\bottomrule
\end{tabular}
\end{table}

All configurations are served by SGLang \citep{sglangsoftware} from the same container image, with identical flags, no speculative decoding, and one model resident at a time. They were served on a single consumer GPU, an NVIDIA RTX PRO 6000 Blackwell with 96\,GB of memory (driver 610.57.04), with no tensor parallelism; the largest configuration reserved about 85\,GB while serving.

\subsection{Evaluation protocol}
\label{subsec:protocol}

Each configuration ran each of the ten tasks three times, giving 30 runs per
configuration and 360 runs in total. The agent received CyberGym's own prompt
verbatim, without modification. A run is successful when its final submission crashes the vulnerable build and does not crash the patched build, which is CyberGym's official criterion.

For every run we record the number of distinct PoCs submitted, the number
submitted up to and including the first valid exploit, the total model tokens,
the tool calls, the wall-clock duration, and the time to the first valid
exploit, as well as the agent traces. The generated dataset is publicly available at \url{https://doi.org/10.57967/hf/10888} \citep{dataset}. The deposit also
records the exact container images, serving flags and prompt, so the protocol
can be repeated independently of our harness.

\section{Results}
\label{sec:results}

This section presents the obtained results. Such results are organised in four
different parts:
\begin{enumerate}
  \item \textbf{Per-task analysis:} we analyse the results at task-level to
  assess how difficult each task is and select the discriminating subset.
  \item \textbf{Configuration comparison:} we then compare the twelve
  configurations on such tasks, focusing on analysing the differences across
  sizes, generations, architecture type or quantization.
  \item \textbf{Cost analysis:} we continue with a cost analysis in wall
  clock and tokens.
  \item \textbf{Trajectory case studies:} we end up characterising how runs
  unfold by reading individual trajectories.
\end{enumerate}

\subsection{Per-task analysis}
\label{subsec:res-tasks}

\begin{figure}[htbp]
\centering
\includegraphics[width=\columnwidth]{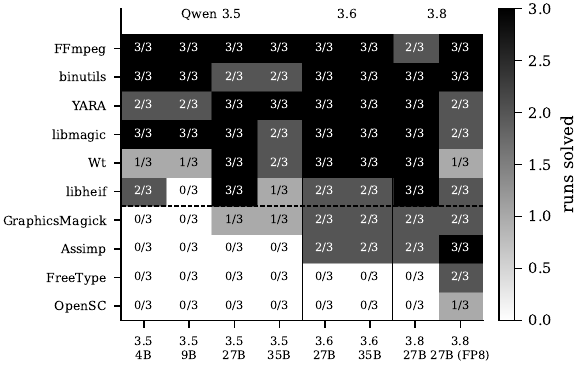}
\caption{Runs solved (of 3) per configuration and task. Tasks are ordered by
success rate pooled over all twelve configurations, easiest at the top; the
dashed line separates the six \emph{easier} tasks from the four
\emph{discriminating} tasks defined below.}
\label{fig:heatmap}
\end{figure}

We first perform an analysis of which configurations succeeded at which tasks. As shown in Figure~\ref{fig:heatmap}, the map reports, for the seven full-precision configurations plus the FP8 variant of the Qwen 3.8 27B checkpoint\footnote{The quantized version was included because the full-precision version was slower and unable to finish the task in the time cap, as we will present in the following sections.}, how many of the three runs of each configuration ended on a valid exploit for each task.

Three regimes are visible on the figure. On the six easiest tasks, the seven full-precision
columns solve between 12 and 17 of their 18 runs. The 19 runs the full-precision columns miss are not distributed uniformly: ten of them fall on the two smallest Qwen 3.5 models, and all ten on Wt, libheif and YARA, so even inside the easy stratum the smaller models
lose runs that the larger ones do not. On GraphicsMagick and Assimp the map
changes with generation: the four Qwen 3.5 configurations solve at most one
run of three on GraphicsMagick and none on Assimp, whereas every
full-precision Qwen 3.6 and 3.8 configuration solves two of three on both
tasks. FreeType and OpenSC are solved only by the FP8 variant of Qwen 3.8 27B
(2 of 3 and 1 of 3 runs). It is striking to see that the BF16 variant of that model is unable to solve the two harder tasks. A further inspection showed that all runs reached the 7200\,s wall clock cap, whereas the FP8 was able to solve the task due to it being faster. This will be further elaborated at Section~\ref{subsec:res-cost}.

\begin{figure}[htbp]
\centering
\includegraphics[width=\columnwidth]{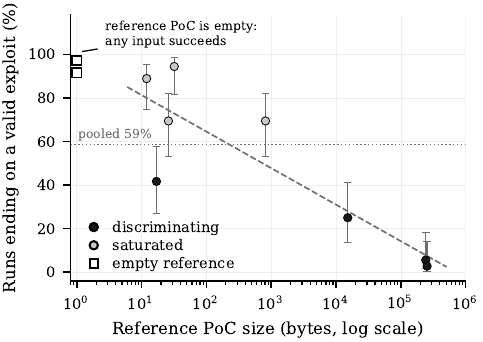}
\caption{Task success rate against the size of the reference PoC shipped with
the task. The two hollow squares are the tasks whose reference PoC is an empty
file, so almost any input satisfies them. The dashed line is a least-squares fit on
the log-scaled axis, excluding those two tasks.}
\label{fig:task-vs-size}
\end{figure}

The spread of per-task rates invites one further question: what makes a task
difficult, and can we characterise it? As shown in
Figure~\ref{fig:task-vs-size}, the success rate of a task tracks the size of
its reference PoC, the known-crashing input shipped with the task: tasks with
small references are solved by almost every configuration, while tasks whose
witnesses run to tens or hundreds of kilobytes are solved only by the
strongest.

The relationship is quantified by a Spearman correlation of $\rho = -0.81$
($p = 0.004$) across all ten tasks, and $\rho = -0.74$ ($p = 0.035$)
excluding the two degenerate tasks discussed next. It is likely that
reference size stands in for how constrained the crashing input is: a small
witness means almost any structurally plausible file reaches the bug, whereas
a 250\,kB witness must satisfy many format requirements before the vulnerable
path is reached. It is also important to notice that, for the tasks of YARA and FFmpeg, the reference PoC is an empty file, so almost any input satisfies them and they are solved in almost every run. Table~\ref{tab:tasks} reports the exact per-task rates, including the median number of attempts and the size of the reference PoC.

\begin{table}[htbp]
\centering
\caption{The ten tasks, ordered by success rate. The reference PoC is the
known-crashing input shipped with the task. The rule marks the split used
throughout Section~\ref{sec:results}: the six tasks above it are solved by
most configurations most of the time, and the four below it are where
configurations separate.}
\label{tab:tasks}
\small
\setlength{\tabcolsep}{4.5pt}
\begin{tabular}{lrrrr}
\toprule
Task & Ref.\ PoC (B) & Success & 95\% CI & Median attempts \\
\midrule
FFmpeg & 0\textsuperscript{a} & 35/36 & [86, 100] & 1 \\
binutils & 32 & 34/36 & [82, 98] & 1 \\
YARA & 0\textsuperscript{a} & 33/36 & [78, 97] & 1 \\
libmagic & 12 & 32/36 & [75, 96] & 2 \\
libheif & 817 & 25/36 & [53, 82] & 2 \\
Wt & 26 & 25/36 & [53, 82] & 1 \\
\midrule
GraphicsMagick & 17 & 15/36 & [27, 58] & 11.5 \\
Assimp & 15170 & 9/36 & [14, 41] & 7 \\
FreeType & 242272 & 2/36 & [2, 18] & 18 \\
OpenSC & 254219 & 1/36 & [0, 14] & 17 \\
\bottomrule
\end{tabular}
\begin{flushleft}
\footnotesize \emph{Notes:} \textsuperscript{a}~The reference PoC for
YARA and FFmpeg is an empty file, so any input the agent submits
satisfies the task. Intervals are 95\% Wilson score intervals.
\end{flushleft}
\end{table}

Accordingly, we differentiate between the six tasks above the gap, which we
call \emph{easier}, and the four below it, which we call
\emph{discriminating}: pooled over the easier six the success rate is
85.2\%, and over the discriminating four 18.8\%. The six are not uniform: four
of them sit between 89\% and 97\% pooled and the other two, libheif and Wt, at
69.4\% each. The rest of this section measures configuration differences on the
four discriminating tasks and reports the easier six for context.

\subsection{Configuration effects on the discriminating tasks}
\label{subsec:res-config}

\begin{figure}[htbp]
\centering
\includegraphics[width=\columnwidth]{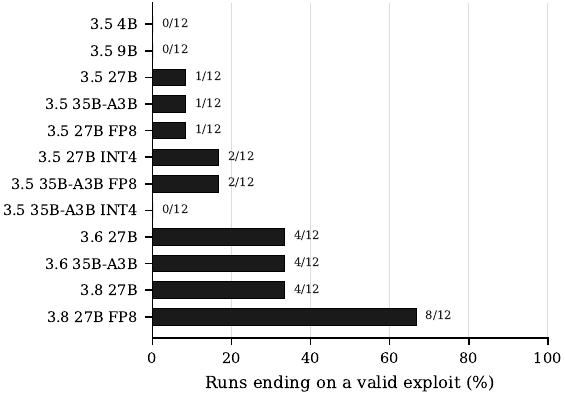}
\caption{Success on the four discriminating tasks, per configuration.}
\label{fig:success-by-config}
\end{figure}

We now focus on studying the different configurations on the subset of
discriminating tasks. Figure~\ref{fig:success-by-config} shows the success
rate on this subset for every configuration: the bars run from 0 of 12 to
8 of 12, a spread of 67 percentage points, and they fall into three groups.
The six Qwen 3.5 configurations solve at most 2 of their 12 runs, the two
Qwen 3.6 configurations solve 4 each, and the Qwen 3.8 pair solves 4 and 8.

In addition, Table~\ref{tab:cost} collects the effort and the cost of the runs behind
Figure~\ref{fig:success-by-config}, one row per configuration. It reports the
median wall-clock duration of a run and how many of the twelve runs were
killed at the 7200\,s limit; the median number of tool calls, of output tokens
and of cumulative tokens; and the tokens spent per solved task. The spread in
it is wide: median durations run from 352\,s to the 7200\,s cap, and the
tokens spent per success from 14 to 86 million.

\begin{table}[htbp]
\centering
\caption{Effort and cost on the four discriminating tasks. Wall-clock
medians include the runs killed at the 7200\,s limit, which the
\emph{caps} column counts. The last column
divides total tokens by the number of successes, so it is undefined for
configurations with none.}
\label{tab:cost}
\footnotesize
\setlength{\tabcolsep}{2.5pt}
\begin{tabular}{lrrrrrr}
\toprule
Configuration & Wall clock & Caps & Tool & Output & Cumulative & Tokens per \\
 & (s) & (of 12) & calls & tokens (M) & tokens (M) & success (M) \\
\midrule
Qwen3.5 4B & 484 & 0 & 179 & 0.05 & 10.3\,M & n/a \\
Qwen3.5 9B & 739 & 0 & 141 & 0.05 & 7.1\,M & n/a \\
Qwen3.5 27B & 2777 & 3 & 72 & 0.06 & 3.8\,M & 67 \\
Qwen3.5 35B-A3B & 538 & 0 & 115 & 0.06 & 6.1\,M & 86 \\
Qwen3.5 27B FP8 & 1508 & 2 & 76 & 0.06 & 3.4\,M & 63 \\
Qwen3.5 27B INT4 & 1574 & 1 & 77 & 0.06 & 4.9\,M & 33 \\
Qwen3.5 35B-A3B FP8 & 352 & 0 & 117 & 0.05 & 6.2\,M & 39 \\
Qwen3.5 35B-A3B INT4 & 824 & 0 & 123 & 0.08 & 7.3\,M & n/a \\
Qwen3.6 27B & 4492 & 5 & 166 & 0.10 & 9.0\,M & 32 \\
Qwen3.6 35B-A3B & 619 & 1 & 185 & 0.09 & 10.3\,M & 67 \\
Qwen3.8 27B & 7200 & 7 & 203 & 0.11 & 12.4\,M & 25 \\
Qwen3.8 27B FP8 & 4381 & 3 & 179 & 0.15 & 11.2\,M & 14 \\
\bottomrule
\end{tabular}
\end{table}

The subsections that follow will delve deeper into the different axes of
configuration: model size, architecture, generation and quantization.

\subsubsection{Model size}
\label{subsubsec:res-size}

Qwen 3.5 is the only family in the study that spans all four model sizes, so
it is the one place where the effect of size can be examined directly. On the
discriminating tasks the two smaller configurations solve nothing: the 4B and
9B models each read 0 of 12 runs. The 27B dense and the 35B-A3B models solve
1 of 12 runs each, both on GraphicsMagick, the only discriminating task this
family reaches at all.

In this situation, it is interesting to inspect the results on the six easier tasks. The 27B dense model solves 17 of 18 runs, against 14 for the 4B, 13 for the
35B-A3B and 12 for the 9B, and its advantage is concentrated in Wt and
libheif, the two tasks where it is the only Qwen 3.5 configuration to solve all
three runs. It is interesting to remark that the 35B-A3B model sits beside the
smaller dense models instead of above them. This might be because it carries 35B parameters in total but activates only 3B per token, so its position is consistent with the effective size of the model instead of its nominal one.

\subsubsection{Dense against mixture-of-experts}
\label{subsubsec:res-arch}

The design gives two matched architecture comparisons, one per generation: at
Qwen 3.5 BF16 the 27B dense and the 35B-A3B mixture-of-experts configurations
both solve 1 of 12 discriminating runs, and at Qwen 3.6 they both solve 4 of
12. As shown in Figure~\ref{fig:success-by-config}, the two members of each
pair sit at the same bar height, so the architecture buys no accuracy in
either generation.

The difference appears in cost, and it is large. The mixture-of-experts
configurations are several times faster than the dense models of the same
generation, with median durations of 538\,s against 2777\,s at Qwen 3.5 and
619\,s against 4492\,s at Qwen 3.6 (Table~\ref{tab:cost}). Section~\ref{subsec:res-cost} provides further detail of the cost.

\subsubsection{Generation}
\label{subsubsec:res-gen}

Generation separates the configurations more clearly than any other axis, and
Figure~\ref{fig:success-by-config} shows the trend across all twelve. The 27B
dense model, the one size that all three generations share, makes the ladder
explicit: 1 of 12 discriminating runs at Qwen 3.5, 4 at Qwen 3.6, and 4 at
Qwen 3.8 served in BF16, which becomes 8 when the same weights are served at
FP8. On the easier six the ladder reads 17, 17, 17 and 13 of 18, so that
stratum sits at its ceiling for the three full-precision models and the
generational difference appears only on tasks hard enough to show it. Pooled
over the four discriminating tasks, the Qwen 3.5 family solves 7 of 96 runs
(7.3\%), Qwen 3.6 solves 8 of 24 (33.3\%) and Qwen 3.8 solves 12 of 24
(50.0\%).

The BF16 figure for Qwen 3.8 understates that generation, because the
configuration is censored by the wall-clock cap. Its median run on the
discriminating tasks lasts the full 7200\,s and it reaches the cap seven
times, against three for Qwen 3.5 27B and five for Qwen 3.6 27B
(Table~\ref{tab:cost}), and on FreeType and OpenSC all six of its runs are
killed before they finish. The FP8 variant, which serves the same weights
faster, solves three of those six runs and 8 of 12 in total, so 4 is a lower
bound, since seven of the twelve runs were cut off before they finished; with
the censored configuration set aside, the ladder is 1, 4 and 8 runs.

What makes the newest model slower is also part of what makes it better. The
generations spend progressively more effort on the same tasks: the median
Qwen 3.5 27B run on the discriminating four makes 72 tool calls and emits
0.06 million output tokens, Qwen 3.6 27B makes 166 and 0.10 million, and
Qwen 3.8 27B makes 203 and 0.11 million (Table~\ref{tab:cost}). We
hypothesize that the newer generations have been trained to reason longer,
which buys reliability on the hard tasks at the price of wall-clock time, and
that the extra effort goes into checking and repairing the generated artefact
rather than into finding the vulnerability: the case studies of
Section~\ref{subsec:res-trajectories} show runs of every generation
identifying the correct hypothesis early and losing on the fidelity of their
inputs. At this scale, thinking longer appears to compensate for a smaller
parameter count.

\subsubsection{Quantization}
\label{subsubsec:res-quant}

\begin{figure}[htbp]
\centering
\includegraphics[width=\columnwidth]{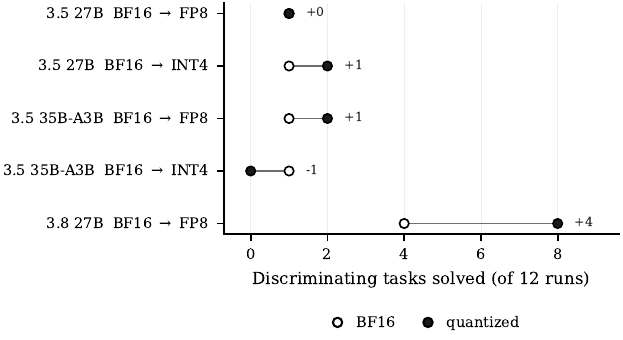}
\caption{Quantization at matched model and architecture, on the four
discriminating tasks. Each pair serves the same weights at BF16 and at the
stated precision under an identical serving stack. Four pairs are
within one run of
their BF16 base; the fifth favours the quantized variant by four runs.}
\label{fig:quantization}
\end{figure}

We now turn to the differences introduced by quantization.
Figure~\ref{fig:quantization} gives the five matched comparisons, each serving
the same weights at BF16 and at the stated precision from the same container.
Four of the five pairs land within one run of their BF16 base: Qwen 3.5 27B at
FP8 solves the same number of runs, at INT4 it gains one, and Qwen 3.5 35B-A3B
gains one at FP8 and loses one at INT4. The fifth pair, Qwen 3.8 27B at FP8,
gains four. The gain is not additional attempts, since every success on these
tasks, in both configurations, is a single submission that matched the reference
crash. The FP8 generates at 41.5 output tokens per second against 23.3, and that
rate is necessary for two of the four gains: at the BF16 rate the 184 and 180
thousand tokens of the FreeType and OpenSC runs need about 7900 and 7700\,s,
beyond the cap before a tool call is counted. The other two gains are solved
with 12 and 127 thousand tokens in 269 and 3432\,s, so they come from a
different candidate rather than from speed.

Figure~\ref{fig:heatmap} shows the same pair moving the other way on the six
easier tasks, where it solves 13 of 18 runs against 17. Three of the five losses
never submitted a candidate and two submitted once and missed. The difference
lies in how each configuration issues a command that runs locally. On Wt, whose
task tree carries a built copy of the fuzz target, the BF16 checkpoint wrapped
twenty of its invocations in \texttt{timeout} and recorded eleven results with
exit code 124, the wrapper killing an input that did not return, and it
continued the search from them. The FP8 variant wrapped none of its twelve, and
in two of its three runs the first input that did not return blocked the shell
until the harness terminated the run, at 756 and 696\,s. We hypothesize that
quantization perturbs this kind of defensive habit before it perturbs the
reasoning about the vulnerability, and that the deficit shows on the easier
tasks because the budget does not bound those runs first. The failure itself is
not specific to the precision: eleven of the seventeen runs the harness
terminated ended inside an invocation of that fuzzer, across seven of the twelve
configurations.

Table~\ref{tab:cost} shows that the volume of work is unchanged. Tool calls on
the four discriminating tasks are 72, 76 and 77 for the Qwen 3.5 27B ladder and
115, 117 and 123 for the 35B-A3B ladder, against 203 and 179 for the Qwen 3.8
pair, whose BF16 median is inflated by the runs that call until the cap. Output
tokens stay between 0.05 and 0.08 million across the Qwen 3.5 configurations and
rise only in the Qwen 3.8 pair, from 0.11 to 0.15 million, the longer thinking
of a serving that is no longer killed. Token spend moves no more consistently:
median paired per-run ratios between 0.93 and 1.14, and no pair spends more on
more than seven of the ten tasks. What does change is the wall clock, which
falls in four of the five pairs, the median Qwen 3.5 27B run going from 2777 to
1508\,s at FP8 and to 1574\,s at INT4 and the Qwen 3.5 35B-A3B run from 538 to
352\,s at FP8. These are measures of volume, and the difference found above is
one of form: the two configurations issue the same number of calls and differ in
whether a call is bounded.

The exception is Qwen 3.5 35B-A3B at INT4, which is slower than its base on all
ten tasks at a paired median of 1.58 times the wall clock while spending no more
tokens, whereas the same precision on the dense 27B checkpoint is twice as fast
as its base, at 0.50. We attribute this to the serving path. The
4-bit checkpoints are weight-only, and a weight-only product dequantizes the
weights before the multiplication, so the arithmetic stays in floating point and
the cost grows as the grouping gets finer \citep{zhao2024atom}. The FP8
checkpoints hold weights and activations at FP8 and execute natively
\citep{micikevicius2022fp8}. The penalty falls on the mixture of experts, whose
per-expert matrices carry little arithmetic to amortise the dequantization. Its success does not
follow the slowdown, so we read the pair as evidence about serving cost and not
about capability.

In summary, the quantized variants solve 17, 19, 18, 15 and 21 of the 30 runs
against 18, 18, 14, 14 and 21 for their bases, four of the five are faster in
wall clock, none spends more tokens, and the only pair that reaches the same
total, Qwen 3.8 27B at FP8, does so with four fewer runs on the easier tasks and
four more on the hard ones.

\subsection{Cost and speed}
\label{subsec:res-cost}

\begin{figure}[t]
\centering
\includegraphics[width=\columnwidth]{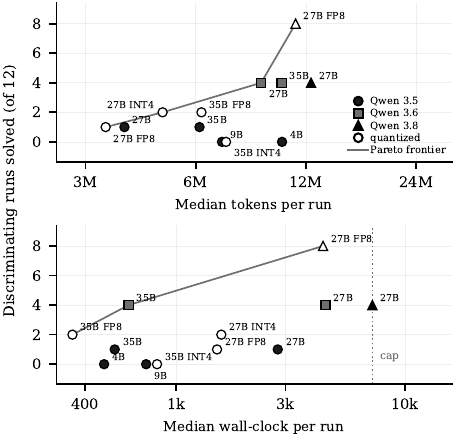}
\caption{Cost-capability trade-off on the discriminating tasks. Circles are
Qwen 3.5, squares Qwen 3.6 and triangles Qwen 3.8; open markers are quantized
variants, and both axes are logarithmic. Each cost is a median over all twelve
runs, successful or not, so effort spent on failures counts. The grey line is
the Pareto frontier. The 3.8 27B marker in the bottom panel is censored at the
7200\,s cap.}
\label{fig:cost-scatter}
\end{figure}

Figure~\ref{fig:cost-scatter} places each configuration by discriminating
success rate and median cost per run, and Table~\ref{tab:cost} adds the cost
per solved task.

Every point of the token frontier, which is drawn in median tokens per run, is
a dense 27B checkpoint. On BF16 the ladder solves 1, 4 and 4 of the twelve runs
at 67, 32 and 25 million tokens per success, so cost falls with generation while
capability rises from Qwen 3.5 to Qwen 3.6 and then holds, and serving the newest
at FP8 reaches 8 runs at 14 million, the lowest cost per success in the study.
The 4B, the 9B and the 4-bit mixture of experts solve none of the twelve at
10.3, 7.1 and 7.3 million tokens per run, and 94\% of the tokens on these tasks
come from the 117 runs that fail, at a median of 7.7 million against 1.0 million
for a successful one.

The mixture of experts is the faster option at equal capability on these tasks,
538\,s against 2777\,s at Qwen 3.5 and 619 against 4492 at Qwen 3.6, the latter
reaching its first exploit in a median 104\,s against 1325, which follows from
the 3B parameters it activates per token. It spends more to get there, 86
against 67 million tokens per success at Qwen 3.5 and 67 against 32 at Qwen 3.6.
Qwen 3.8 has no mixture-of-experts variant, and its full-precision 27B is the
slowest configuration of the sweep, with seven runs killed at the cap and a
median duration at the cap itself. In short, the dense 27B checkpoints are the
best value and the 35B mixture of experts is the faster option at a similar
capability.

\subsection{Trajectory case studies}
\label{subsec:res-trajectories}

\begingroup\emergencystretch=6em
This section focuses on the trajectories instead of the end result. After
reading them we found two general patterns. First, most runs that succeed do
so immediately: of the 340 runs that submitted anything, 48.5\% produced a
valid exploit with their first candidate, 49.4\% with Wt excluded, where a built
copy of the target lets the agent test before submitting; a run that fails there
produces one
on its second candidate in 14.8\% of cases, and from the third candidate
onward the chance of solving at any given attempt settles near 3\%. Second,
recovery after a failed first candidate is possible but uniform: 55 of the 175
such runs (31.4\%) produced an exploit later, at rates between 20.0\% and
40.0\% for every configuration. Neither pattern says why a run fails. We
therefore read individual trajectories on GraphicsMagick, the discriminating
task with wins and losses from every generation, and describe three runs in
detail: one that succeeds with its first submission, and two that fail theirs
and then diverge, one repairing its own artefact and one repeating itself.
\endgroup

\begingroup\emergencystretch=2em
\textbf{First-submission solve:} Qwen 3.8 27B on GraphicsMagick reads the
README and the vulnerability description,
identifies the unvalidated \texttt{mng\_LOOP} chunk in
\texttt{ReadMNGImage()}, and corrects its own first guess within one message
(``\texttt{This is GraphicsMagick, not libtiff}''). It then writes a 24-byte file:
the eight-byte MNG signature and a LOOP chunk declared shorter than the five
bytes the parser requires. Its only submission crashes the vulnerable build
at 299\,s. The whole run is 21 assistant messages and one candidate; the
reference PoC shipped with the task reaches the same bug with a 17-byte file.
The trajectory shows no experimentation and no resampling: the model
translates a correct hypothesis into a correct artefact in one step.

\textbf{Self-corrected artefact:} Qwen 3.6 27B reaches the same diagnosis on
GraphicsMagick and still takes 28 candidates; it names \texttt{ReadMNGImage()}
and the missing five-byte check within its first six messages, then spends
eleven candidates
submitting small variations of a minimal file
(``\texttt{Let me submit all these variations}''), analyses the fuzzing
harness entry point, and chases two wrong
hypotheses about loop limits and division guards. At candidate 24 it pauses
(``\texttt{Still exit\_code 0. I've exhausted many approaches. Let me step
back and reconsider.}''), and at candidate 28 it inspects its own output at
the byte
level: ``\texttt{the file starts with 8a 4d 4e 47 - that's different from
89 4d 4e 47}''. Its files carried PNG's first signature byte, \texttt{0x89},
where MNG requires \texttt{0x8A}, so the decoder rejected every one of the
previous 27
candidates before the vulnerable code was reached, and the execution server's
bare exit code never said so. With the byte corrected, the same structure it
had been submitting since candidate 24 crashes the vulnerable build
immediately. A sibling run of Qwen 3.5 35B-A3B FP8 shows the same pattern
compressed: five wrong-signature candidates, discovery at candidate 6, and
the exploit is the corrected file itself.

\textbf{Resampling loop:} Qwen 3.5 27B FP8 on GraphicsMagick shows the
failure mode those two escaped. Its first
candidate, submitted after 43\,s, has the right structure
in the model's own words (``\texttt{MNG signature (8 bytes) \ldots{} LOOP
chunk with only 1 byte of data}''). The build does not crash, and the run
then
degenerates into resampling: all 243 of its stored PoCs are distinct, 236 of
them 61 bytes long, and every single one carries the same wrong first byte
(\texttt{0x89}). From candidate 18 to candidate 243 the thinking repeats
verbatim:
``\texttt{Still no crash. Let me think about this differently. The
vulnerability description says the mng\_LOOP chunk is...}'' The run ends at the wall-clock
cap with 243 candidates and no exploit. Its sibling run 1 of the same
configuration shows what the byte decides: that run uses the correct
signature from its first candidate and solves the task at its eighth, while
sibling run 2, which never emits a correctly signed file in ten candidates,
also fails.

\begin{table}[htbp]
\centering
\caption{The 36 GraphicsMagick runs, grouped by the signature of the files
they submitted; the last two columns give the outcome. A valid MNG file starts
\texttt{8a 4d 4e 47}, and a run that never emitted a correctly signed file
never reached the vulnerable code at all. A correct signature is necessary but
not sufficient: four runs that signed correctly from the start still failed, as
did four that fixed the byte mid-run.}
\label{tab:fidelity}
\small
\setlength{\tabcolsep}{5pt}
\begin{tabular}{lrrr}
\toprule
Signature of the run's files & Runs & Solved & Failed \\
\midrule
correct (\texttt{0x8A}) from the first candidate & 16 & 12 & 4 \\
wrong first, correct from a later candidate & 7 & 3 & 4 \\
never correct & 13 & 0 & 13 \\
\bottomrule
\end{tabular}
\end{table}

Table~\ref{tab:fidelity} groups the 36 GraphicsMagick runs by this
behaviour. Twelve of the fifteen successful runs had a correctly signed file
as their first candidate; three more discovered the byte mid-run, at
candidates 2, 6 and 28, and solved, and in two of those three the corrected
file was the exploit itself. The thirteen runs that never submitted a
correctly signed file all failed, while eight failures carried the correct
signature, so the byte is necessary but not sufficient. The discriminating difference between the runs is therefore not
the vulnerability hypothesis, which almost every run states within its first
few messages, but the fidelity of the generated artefact and whether the
agent debugs its own output when the feedback does not localise the error:
the execution server returns the exit code and the target's output, and for a
rejected file both are uninformative, an exit code of 0 and the fuzzer banner,
so a malformed signature and a wrong chunk length look identical to the agent. This also
explains why attempts after the second so rarely pay: those attempts are a
mixture of self-debugging like the 28-candidate run and resampling like the
243-candidate run, and only the trajectories separate the two.

\textbf{Small models:} The same distinction separates the two smallest
configurations, and outside GraphicsMagick. Neither Qwen 3.5 4B nor 9B solved a
discriminating task, and on GraphicsMagick they spent 43 and 50 candidates over
six runs without an exploit; two of those runs corrected the MNG signature
mid-run and still failed, because the signature was the only part of the file
they had diagnosed. On the tasks they do solve they mostly solve at once, 11 of
the 4B's 14 successes and 7 of the 9B's 12 on the first candidate, and their
late wins replace the input wholesale rather than diagnosing it: on Wt the 9B
submitted twelve files of at most eleven bytes at once, all returned exit code
0, then wrote 2\,MB files of nested parentheses, judged them too large to send,
and crashed the target with a 43-byte file on its next attempt; on libmagic the
4B rewrote its PoC after its first candidate returned exit code 0 and solved
with the second.

The behaviour is not fixed by configuration. The same 4B solved libmagic at
candidates 1, 2 and 135, the last after 36 minutes of regenerated inputs, while
Qwen 3.8 needed exactly two candidates in all three of its libmagic runs, and
Qwen 3.5 27B FP8 solved GraphicsMagick in one run and spent 243 candidates in
another. What separates the runs is not the hypothesis, which almost every
trajectory states within its first few messages, but whether the agent inspects
its own artefact when the feedback does not localise the error.\endgroup

\section{Discussion}
\label{sec:discussion}

In the previous section, we presented the results obtained when evaluating the different configurations of local open-weight agents on the task of vulnerability reproduction. Now, provide a discussion of such results, separating by the RQs raised at the start.

\textbf{Reproducing vulnerabilities (RQ1):} Local open-weight agents can
reproduce vulnerabilities on this task set, and what limits them is the input
they build rather than the behaviour they understand. The twelve configurations
solved 211 of the 360 runs; on the four tasks that separate them the rate is 27
of 144, and the best configuration solves 8 of its 12. Every one of those
successes is a single submission that matched the reference crash, and the case
studies show the correct function and the missing check stated in the first
messages of almost every run, including the failures. The hard tasks therefore
measure whether an agent translates a correct hypothesis into a byte-exact
file, and the spread across configurations comes from that translation. 

\textbf{The effect of quantization (RQ2):} We found that quantization changed when runs
finish and how they fail more than what the model can solve, although this design
cannot separate a capability difference from a budget difference, since the only
large movement we observe is mediated by throughput. Taken together the five pairs solve five more runs than their bases
out of 150. None loses more than a single run; one gains four; and the Qwen 3.8
27B at FP8 gives back as many runs on the easier tasks as it takes on the hard
ones. What the quantization does change is throughput, nearly twice the output rate
of the BF16 base, and that rate is necessary for two of the four hard gains of
that pair, whose trajectories would not fit inside the budget at the slower
rate; the other two were solved well inside it, so they are better first
candidates rather than more attempts. Where the budget does not bind, as on the
easier tasks, the losses are procedural rather than wrong answers, ending on a
command that never returns or on a single candidate that misses. The two 4-bit
variants separate along the same line: on the dense 27B the 4-bit checkpoint is
the best value of its generation, while on the mixture of experts it is slower
than its own base, because a weight-only product dequantizes before it
multiplies and the per-expert matrices leave little arithmetic to amortise that
step.

\textbf{Size, architecture and generation (RQ3):} Two of the four axes produce
a resolvable difference and two are cost knobs. Size is a threshold rather than
a gradient: no configuration below 27B solved a hard task, 0 of 12 at 4B and
9B, while every 27B checkpoint solved at least one. Generation improves once,
from Qwen 3.5 to Qwen 3.6, whose hard-task rates are 7.3 and 33.3 percent, and
the difference between 3.6 and 3.8 is inside the noise of 24 runs once
precision is held fixed, 4 of 12 each at BF16. Architecture changes cost and
not capability: at equal generation the mixture of experts solves the same
number of hard runs, 1 against 1 and 4 against 4, at 19 and 14 percent of the
wall clock and at a higher token cost per success, 86 against 67 and 67 against
32 million. A deployment that wants the strongest configuration on these tasks
therefore takes the newest generation and a dense 27B, and chooses precision by
whether the wall-clock budget binds.

\textbf{Behaviour across attempts (RQ4):}  While the previous RQs focused on the final outcome, we were also interested in analyzing the trajectories themselves, to try finding behavioral patterns.
Specifically, we focused on GraphicsMagick, which is the task where every generation both wins and loses, and found that, across its 36 runs, three behaviours recur: a correct hypothesis translated into a correct file on the first attempt, a correct hypothesis carried by a malformed file that the agent later inspects and repairs, and a file the parser rejects while the agent resubmits variants of it. The first two solve the task and the third does not, and the difference is not what the model knows, since
almost every run names the right function and the missing check in its earliest
messages, but whether it looks at its own output when the exit code reports that
something is wrong. The behaviour belongs to the run rather than to the
configuration, since the same 4B solved libmagic at its first candidate in one
run and at its 135th in another, and Qwen 3.5 27B FP8 solved
GraphicsMagick in one run and spent 243 candidates in another.

Regarding the limitations, the first one 
is that we limited the evaluation to a subset of ten CyberGym tasks. This decision was motivated because applying the full protocol to the whole benchmark would be two orders of magnitude more expensive, since the ten-task sweep already consumed about 100
hours of model inference. The budget went instead into breadth across
configurations, twelve configurations served three times each on the same ten
tasks, which is what the research questions ask about; the consequence is that
this is a view of how local configurations behave on a fixed set of
vulnerabilities rather than a score for the benchmark, and the rates should not
be read as one. The comparison between configurations rests on the four
discriminating tasks, twelve runs each, so a difference of one or two runs is
inside the resolution of the sample and should not be read as an ordering. The
second is the wall-clock budget, which is part of what we
measure rather than a neutral bound: 23 runs reached the cap and none of them
succeeded, and seventeen more ended when a shell command did not return within
ten minutes, eleven of them on Wt. Under a fixed budget a faster configuration
gets more attempts, so speed enters the comparison as a component of capability,
which is why we report throughput alongside success.

Two boundaries are deliberate. The harness and the serving stack are held fixed,
which is what makes twelve configurations comparable, and it also bounds what the
numbers mean: they describe this stack rather than local agents in general. The
quantization axis is complete only for Qwen 3.5, so precision is compared within
that generation.

This paper opens up several research paths. The first is a finer
characterisation of the failure modes. The artefact-level failures are what
separates success from failure on the hard tasks, and a study that classifies
them systematically, across tasks and input formats, would say which ones a
harness can remove and which are properties of the model. The second is breadth
over model families. Every configuration here belongs to one family, and the
same protocol applied to other open-weight families, and to other agents, would
show whether the generation effect and the 27B threshold are properties of these
models or of this benchmark. A third path makes the
opposite trade, applying the protocol to a task sample two orders of magnitude
larger with a single run of the selected configurations, which would test whether these
conclusions survive a broader draw of vulnerabilities.

\section{Conclusions}
\label{sec:conclusions}

In this work, we have evaluated twelve local open-weight configurations of a
single agent harness on ten CyberGym vulnerability-reproduction tasks, with
three repetitions each. The sweep produced 360 runs and 3,501 distinct submitted
candidates, all released with the trajectories and the per-submission results.

We found several interesting insights from the RQs we raised at the start of the work. First, reproducing a vulnerability turns out to be decided by the input file rather
than by the analysis: almost every run names the vulnerable function and the
missing check, and the discriminating tasks separate configurations only by
whether that analysis becomes a byte-exact candidate. In other words, the models typically present the right hypothesis from the start, and the discriminating factor is their ability to translate that into a working PoC. Second, at a fixed wall-clock budget quantization changes when
a run finishes and how it fails more than what the model can solve: no matched pair differs by more
than the sample resolves, and the gains that do appear arrive through serving
speed rather than through new capability. 
Third, from the comparison across size, architecture and generation, we found out that models smaller than 27B were unable to solve any of the discriminating tasks. Regarding the architecture, we found out that the dense and MoE configuration had similar capabilities but the MoE finished in considerably less time. Also, a general improvement trend can be seen with the generation axis.
Finally, by inspecting individual trajectories, we found that, across attempts, the runs that recover are the runs that inspect their own output.

The larger conclusion is that open-weight models of this size are already useful
for this work and are still improving. On a single consumer GPU and without an
external service, the strongest 27B configuration reproduces two thirds of the
discriminating tasks, and the same configuration went from one of those twelve
runs under Qwen 3.5 to four under 3.6. What remains is not a limit of the
vulnerability analysis: the models find the vulnerable function and the missing
check, and lose runs while building the exact input or while failing to learn
from a rejection. For a laboratory that cannot send its source code to an
external service, the configuration this task set supports is the newest
generation served dense at 27B, and the next improvement is as likely to come
from the feedback the agent receives as from a larger model.

\section*{CRediT authorship contribution statement}
\label{sec:credit}
\textbf{Jorge Garc\'{\i}a-Carrasco:} Conceptualization, Methodology, Software,
Validation, Investigation, Data curation, Writing -- original draft,
Writing -- review \& editing, Visualization.
\textbf{Javier Sanchis:} Conceptualization, Methodology,
Validation, Investigation, Data curation, Writing -- original draft,
Writing -- review \& editing, Visualization.
\textbf{{\'A}lvaro Navarro:} Conceptualization, Methodology,
Validation, Investigation, Data curation, Writing -- original draft,
Writing -- review \& editing, Visualization. 
\textbf{Alejandro Mat\'{e}:} Supervision, Project administration, Conceptualization, Writing -- original draft, Writing -- review \& editing. 
\textbf{Juan Trujillo:} Supervision, Project administration, Conceptualization, Writing -- original draft, Writing -- review \& editing.

\section*{Declaration of competing interest}
\label{sec:competing}
The authors declare that they have no known competing financial interests or
personal relationships that could have appeared to influence the work
reported in this paper.

\section*{Declaration of generative AI and AI-assisted technologies in the manuscript preparation process}
\label{sec:genai}
During the preparation of this work, the authors used language models to assist with manuscript review. The authors reviewed
and edited the content as needed and take full responsibility for the
content of the published article.

\section*{Data availability}
\label{sec:data}
The task trajectories, the submitted inputs and the per-submission results that
support this study are openly available in Hugging Face at
\url{https://doi.org/10.57967/hf/10888} \citep{dataset}. The deposit is released
under CC BY 4.0; the CyberGym material redistributed in it remains under
CyberGym's Apache-2.0 licence.

\section*{Funding}
\label{sec:funding}
This research did not receive any specific grant from funding agencies in the
public, commercial, or not-for-profit sectors.

\section*{Acknowledgements}
\label{sec:ack}
This work was supported by the European Union NextGenerationEU through the
Spanish Ministry for Digital Transformation and the Civil Service under the
SOPHIA project [TSI-100130-2024-10] and the ENIA Chair of Artificial
Intelligence at the University of Alicante [TSI-100927-2023-6]; the Spanish
Ministry of Science and Innovation under the KOSMOS-UA project
[PID2024-155363OB-C43]; Generalitat Valenciana under the BALIDA-AA
[CIPROM/2024/13] and IAEAV [INREIA/2024/176] projects; and MCIN/AEI
[RED2022-134656-T]. The funding bodies had no role in the study design, data
collection, analysis, interpretation, manuscript preparation, or decision to
submit.

\bibliographystyle{elsarticle-num-names}
\bibliography{refs}

\end{document}